\documentclass[aps,twocolumn,showpacs]{revtex4}
\usepackage{amsmath,amssymb,amsfonts}
\usepackage{graphicx,graphics,psfig}
\usepackage{color}
\allowdisplaybreaks

\newcommand{\be}{\begin{equation}}
\newcommand{\ee}{\end{equation}}
\newcommand{\bea}{\begin{eqnarray}}
\newcommand{\eea}{\end{eqnarray}}
\newcommand{\bm}{\bibitem}

\newcommand{\gm}{\gamma}

\newcommand{\ep}{\epsilon}
\newcommand{\de}{\delta}
\newcommand{\De}{\Delta}
\newcommand{\om}{\omega}
\newcommand{\Om}{\Omega}
\newcommand{\tht}{\theta}

\newcommand{\Lm}{\Lambda}
\newcommand{\sg}{\sigma}
\newcommand{\Sg}{\Sigma}
\newcommand{\oD}{\overline{D}}
\newcommand{\oS}{\overline{S}}

\newcommand{\psb}{\overline\psi}
\newcommand{\Psb}{\overline\Psi}
\newcommand{\wt}{\widetilde}
\newcommand{\vk}{\vec k}
\newcommand{\vp}{\vec p}
\newcommand{\vq}{\vec q}

\newcommand{\lgl}{\langle}
\newcommand{\rgl}{\rangle}
\newcommand{\bL}{\boldsymbol{\Lambda}}
\newcommand{\bom}{\boldsymbol{\Omega}}
\newcommand{\bD}{\boldsymbol{D}}
\newcommand{\bU}{\boldsymbol{U}}
\newcommand{\bV}{\boldsymbol{V}}
\newcommand{\bsg}{\boldsymbol{\Sigma}}
\newcommand{\bS}{\boldsymbol{S}}
\newcommand{\ps}{p \!\!\! /}
\newcommand{\rw}{\rightarrow}
\begin{document}

\setcounter{page}{1}

\title{Real-time propagators at finite temperature and chemical potential}

\author{S. \surname{Mallik}}
\email{mallik@theory.saha.ernet.in}
\affiliation{Theory Division, Saha Institute of Nuclear Physics, 1/AF 
 Bidhannagar, Kolkata 700064, India}  
\author{Sourav \surname{Sarkar}}
\email{sourav@veccal.ernet.in}
\affiliation{Variable Energy Cyclotron Centre, 1/AF, Bidhannagar, 
Kolkata, 700064, India}

\date{\today}

\begin{abstract}
We derive a form of spectral representations for all bosonic and 
fermionic propagators in the real-time formulation of field theory at finite 
temperature and chemical potential. Besides being simple and  symmetrical 
between the bosonic and the fermionic types, these representations depend 
explicitly on analytic functions only. This last property allows a simple 
evaluation of loop integrals in the energy variables over propagators in this 
form, even in presence of chemical potentials, which is not possible over
their conventional form.
\end{abstract}  


\maketitle
\section{Introduction}
An advantage of the real-time formulation of field theory in medium over its
imaginary-time counterpart is its close resemblance with the field theory in
vacuum \cite{Semenoff,Niemi,Kobes1,Landsmann}. A matrix propagator in this 
formulation is given, roughly speaking, by the vacuum propagator and another piece 
representing the medium. Also one deals here with an energy integral instead of 
a frequency sum. Thus, even though its matrix structure is looked upon as a
source of complication, the real-time formulation has found many
applications to study the properties of hadrons propagating through a medium 
\cite{Leutwyler,Schenk,Toublan,Nieves,Pisarski,Mallik,MS}.
 
However, a difficulty arises with the conventional form of propagators in
presence of one or more chemical potentials. Such propagators have theta-functions 
in the energy variable separating the contributions of particles and antiparticles 
in the medium. Then, unlike the case in vacuum, the energy integral in a loop
contribution cannot be carried out by the simple contour method
~\cite{Kobes2,Comment1}.

In this work we rewrite all the real-time propagators in terms of their 
K\"{a}llen-Lehmann spectral representations \cite{Kallen,Lehmann} in a suitable 
form. Besides the spectral densities, these representations depend only on analytic 
functions, even in presence of chemical potentials, allowing contour evaluation of 
integrals in loop energies. The integrations over the (known) spectral densities can 
then be performed, bringing out the different distribution functions for particles 
and antiparticles, when chemical potentials are present.

As expected, our representations are also valid for complete propagators, in fact, 
for any two-point correlation functions. These are found to be only of two types, 
depending on the bosonic or the fermionic character of the operators in the 
two-point functions. The two types of matrix propagators are rather simple
and enjoy a high symmetry between them.

We begin with representations for the three-dimensional (spatial) Fourier
transforms of the two-point functions. After selecting the time contour, we
carry out the remaining Fourier transform with respect to time to get the 
desired representation. These, of course, reproduce the conventional form of
free propagators, once the integrations over their known spectral densities
are carried out.

In Sec 2 we consider the spectral representations for the spatial Fourier
transform of the {\it free} propagators. In Sec 3 we show that the same form is 
also valid for the {\it complete} propagators. Then in Sec 4 we obtain the four 
dimensional transforms. In Sec 5 we work out an example of a loop integral with 
bosonic and fermionic propagators. Our discussion is contained in Sec 6. 

\section{Free Theory}

Let us denote the ensemble average of an operator $O$ by $\lgl O\rgl$,
\be
\lgl O\rgl=Tr\,[e^{-\beta(H-\mu N)}O]/Z,~~~~~~Z=Tr\,[e^{-\beta(H-\mu N)}]\,,
\ee
where $H$ and $N$ denote the Hamiltonian and the number operator of the
system under consideration, maintained at temperature $1/\beta$ and chemical
potential $\mu$.

We begin with the {\it free} propagators for a scalar field $\phi (x)$ (with no
chemical potential) and a (Dirac) spinor field $\psi (x)$ (with chemical
potential $\mu$), 
\bea
&& D^{(0)}(x-x')=i\lgl T_c\phi(x)\phi(x')\rgl,\nonumber\\
&& S^{(0)}(x-x')=i\lgl T_c\psi(x)\psb (x')\rgl   
\eea
where $x^\mu=(\tau,\vec x)$. Here $\tau$ and $\tau'$ are any two 'times' on a
contour in the complex time plane. The ensemble average with the Boltzmann 
weight and the consequent analyticity domain,  
$-\beta\le\mathrm{Im}(\tau-\tau')\le 0$, constrains the contour to begin at $-T$, 
say, on the real axis and end at $-T-i\beta$, nowhere moving upwards \cite{Mills}. 
Apart from these requirements, we let the contour be arbitrary at this stage. 
Here $T_c$ (and $\theta_c$ below) denote time ordering (and $\theta$-function) 
with respect to such a contour.

Our form for the propagators in momentum space follow directly from the following 
spectral representations of their spatial Fourier transforms \cite{Mills}, 
\bea
&&\!\!\!\!\!\!\!\!\!\!\!D^{(0)}(\tau-\tau',\vk)=\nonumber\\
&&\!\!\!\!\!\!\!\!\!\!\!i\int_{-\infty}^{+\infty}\frac{dk_0'}{2\pi}\rho^{(0)}(k_0',\vk)
e^{-ik'_0(\tau-\tau')}\{\tht_c(\tau-\tau')+f(k_0')\}
\label{spftd}
\eea
\bea
&&\!\!\!\!\!\!\!\!\!\!\!S^{(0)}(\tau-\tau',\vp)=\nonumber\\
&&\!\!\!\!\!\!\!\!\!\!\!i\int_{-\infty}^{+\infty}\frac{dp_0'}{2\pi}\sg^{(0)}(p_0',\vp)
e^{-ip'_0(\tau-\tau')}\{\tht_c(\tau-\tau')-\wt{f}(p_0')\}
\label{spfts}
\eea
where $\rho^{(0)}$ and $\sg^{(0)}$ are the bosonic and the fermionic spectral 
functions for the free theory,
\be
\rho^{(0)}(k_0,\vec k)=2\pi\ep(k_0)\de(k^2-m_B^2)
\label{frspfn_b}
\ee
\be
\sg^{(0)}(p_0,\vec p)=2\pi\ep(p_0)(p\!\!\!/+m_F)\de(p^2-m_F^2)
\label{frspfn_f}
\ee
and the functions $f$ and $\wt f$ are given by
\be
f(k_0)=\frac{1}{e^{\beta k_0}-1}~~,~~~~~~~~~~
\wt{f}(p_0)=\frac{1}{e^{\beta(p_0-\mu)}+1}
\ee
As $k_0$ and $p_0$ run over the entire real axis, $f$ and ${\wt f}$ cannot
be interpreted as particle distribution functions, which are
\bea
n(k_0)&=&\frac{1}{e^{\beta |k_0|}-1}~~~(\mathrm{boson})\nonumber\\
n_+(p_0)&=&\frac{1}{e^{\beta(|p_0|-\mu)}+1}~~~(\mathrm{fermion})\nonumber\\
n_-(p_0)&=&\frac{1}{e^{\beta(|p_0|+\mu)}+1}~~~(\mathrm{antifermion})
\eea
But they may be readily expressed in terms of the latter ones as
\be
f(k_0)=n\ep(k_0)-\tht(-k_0),~~~~\wt f(p_0)=N_1\ep(p_0) +\tht(-p_0)
\ee
where
\be
N_1(p_0)=n_+\tht(p_0) +n_-\tht(-p_0)
\ee

A simple way to derive the results (3) and (4) is to note that each of the
propagators in medium satisfy the same differential equation as the
corresponding one in vacuum. Taking their spatial Fourier transforms we get
one dimentional equations in time $\tau$. A particular solution to such an
{\it inhomogeneous} equation is given by the one with Feynman boundary condition. 
To get the most general solution, we add to it the two independent (plane wave) 
solutions of the {\it homogeneous} equation with arbitrary coefficients. When these 
coefficients are determined by applying the thermal, so-called Kubo-Martin-Schwinger 
(KMS) boundary condition \cite{Kubo,Martin}, we get the above solutions.
One can, of course, readily verify these solutions by showing that they 
satisfy the respective differential equation as well the boundary condtion.

\section{Interacting theory}
Before we proceed, let us consider complete propagators for interacting theories 
in the Heisenberg picture and obtain the K\"allen-Lehmann representation for their
spatial Fourier transforms. Consider first the full propagator for the 
interacting scalar field $\Phi(x)$,
\be
D(x-x')=i\lgl T_c\Phi(x)\Phi(x')\rgl\,.
\label{dp1}   
\ee
Now it does not satisfy a simple differential equation any more. But we may
analyze the field product by introducing a complete set of eigenstates $|m\rgl$
of the four-momentum operator $P_\mu$ with eigenvalues $(E_m,\vec k_m)$. We
insert this set twice in Eq.~(11), once to evaluate the ensemble trace and then 
to separate the operators in the product. Using translational invariance
to extract the $x$-dependence of the operators and introducing a delta-function 
in the energy variables, the spatial Fourier transform of $\lgl\Phi(x)\Phi(x')\rgl$  
in the $T_c$-product becomes
\be
\int d^3x\, e^{-i\vec k\cdot(\vec x-\vec
x')}\lgl\Phi(x)\Phi(x')\rgl=\int\frac{dk_0'}{2\pi}
e^{-ik_0'(\tau-\tau')}M^+(k_0',\vec k)
\ee
where the spectral density $M^+$ is given by
\bea
\!\!\!\!M^+(k_0,\vec k)=(2\pi)^4\sum_{m,n}&&\!\!\!e^{-\beta
E_m}\de^{(4)}(k+k_m-k_n)\times \nonumber\\
&&\lgl m|\Phi(0)|n\rgl\lgl n|\Phi(0)|m\rgl/Z\,.
\eea
In just the same way we work out the Fourier transform $\lgl\Phi(x')\Phi(x)\rgl$ 
with a spectral density $M^-$, that may be obtained from $M^+$ by changing the
signs of $k_m$ and $k_n$ in the delta-function. The two spectral densities are 
related by the KMS condition
\be 
M^+(k)=e^{\beta k_0}M^-(k)
\ee 
which may be obtained simply by interchanging the dummy indices $m,n$ in one of 
$M^\pm(k)$ and using the energy conserving $\de$-function~\cite{Comment2}.

We now define a third spectral density related to the {\it commutator} of
the field operators,
\be
\rho(k)=M^+(k)-M^-(k)\equiv\int d^4x e^{ik\cdot(x-x')}\lgl
[\Phi(x),\Phi(x')]\rgl
\ee
in terms of which $M^\pm(k)$ can be expressed as
\be
M^+(k)=\{1+f(k_0)\}\rho(k), ~~~~
M^-(k)=f(k_0)\rho(k)
\ee
Using these relations, we see that the three-dimensional Fourier transform of the 
complete propagator is given exactly by Eq.(3) with the free density $\rho^{(0)}$ 
replaced by the interacting one, $\rho$. 

The complete propagator for the interacting spinor field $\Psi (x)$,
\be
S(x-x')=i\lgl T_c\Psi(x)\Psb(x')\rgl
\ee
can be treated in the same way. The three dimensional Fourier transform of
$\lgl\Psi(x)\Psb(x')\rgl$ is given by an equation like (12) with the spectral 
density
\bea
\wt{M}^+(p_0,\vec p)=(2\pi)^4\sum_{m,n}&&e^{-\beta (E_m-\mu N_m)}
\de^{(4)}(p+p_m-p_n)\times\nonumber\\
&&\lgl m|\Psi(0)|n\rgl\lgl n|\Psb(0)|m\rgl/Z\,,
\eea
while that of $\lgl\Psb(x')\Psi(x)\rgl$  is given by a second spectral
density $\wt{M}^-(p_0,\vec p)$, obtained from the first one by interchanging
$\Psi(0)$ and $\Psb(0)$  and changing the signs of $p_m$ and $p_n$. Here
the integers $N_m$ are the eigenvalues of the operator $N$, the fermion
number conservation restricting them to $N_m=N_n-1$.
The two spectral densities are again related by the KMS condition
\be 
\wt{M}^+(p)=e^{\beta (p_0-\mu)}\wt{M}^-(p)\,,
\ee 

Defining a third spectral density in terms of the {\it anti-commutator} of the
field operators,
\be
\sg(p)=\wt{M}^+(p) +\wt{M}^-(p) \equiv\int d^4x e^{ip\cdot(x-x')} \lgl \{\Psi
(x),\overline{\Psi} (x')\}\rgl
\ee
we can express the first two as
\be 
\wt{M}^+(p)=\{1-\wt{f}(p_0)\}\sg (p),~~~~~~~\wt{M}^-(p) =\wt{f}(p_0)\sg(p)
\ee
We thus again get the complete spinor propagator as given by
Eq.(4) with $\sg^{(0)}$ replaced by $\sg$.

It should be noted that in deriving these representations for
the complete scalar and spinor propagators, we have not used any property 
specific to these field operators, except to define their spectral densities
by the commutator and the anticommutator of the respective fields. We thus
conclude that these are the most general representations for the spatial
Fourier transforms of two-point functions of {\it any} bosonic and fermionic 
local operators. Below we construct their real time representations, again without 
requiring any property of the spectral densities at all, so that although we refer 
to them as propagators, they actually remain valid for any two-point function.   

In passing, let us compare the derivation of the K\"{a}llen-Lehmann representation 
for the ensemble average of two time-ordered fields with that for its vacuum
expectation value. In the vacuum case, the intermediate energies are all positive 
and Lorentz invariance makes the spectral densities to depend on $\tht(k_0)$ and 
$k^2(=k_0^2-\vk^2)$ ($\tht(p_o)$ and $ p^2(=p_0^2-\vp^2)$). There is, of course, 
no KMS condition here, but since the commutator (anticommutator) vanishes outside 
the light-cone, the two parts of spectral densities must be equal \cite{Weinberg}. 
In the medium, there is no Lorentz invariance (or  we may have it at the cost of 
introducing a four-velocity vector for the medium). As a result, the vanishing of 
the commutator (anticommutator) does not lead to any simple condition on the spectral
densities.

\section{Form of propagators}
Having established the results for the spatial Fourier transforms of the 
propagators, where we kept their time coordinates on an arbitrary contour in 
the complex time plane, we now choose an appropriate contour to get 
the real time field theory. Our choice is that of Fig.~1, which for $T\to\infty$, 
reduces to two parallel lines, the real line and the one shifted by $-i\beta/2$, 
to be denoted by subscripts $1$ and $2$ respectively \cite{Semenoff}. 
Thus $\tau_1=t,\, \tau_2=t-i\beta/2,\, \tht_c(\tau_1-\tau'_1)=
\tht (t-t'),\, \tht_c (\tau_1-\tau'_2)=0,$ etc. We can now define
the Fourier transform of the four components of a propagator with respect 
to real time. We thus get from Eqs.~(3, 4) the four dimensional Fourier transform 
of the components respectively of the boson and the fermion propagator as 
\be
D_{ab}(k_0,\vec k)=\int_{-\infty}^{+\infty}dt\,e^{ik_0(t-t')}
D(\tau_a-\tau'_b,\vec k),~~~~~a,b=1,2
\ee
\be
S_{ab}(p_0,\vec p)=\int_{-\infty}^{+\infty}dt\,e^{ip_0(t-t')}
S(\tau_a-\tau'_b,\vec p),~~~~~a,b=1,2
\ee

\begin{figure}
\centerline{\psfig{figure=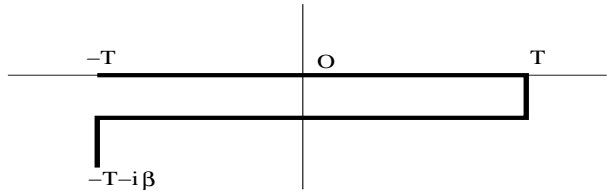,height=2.5cm,width=8cm}}
\caption{Contour in time plane for real time formalism}.
\end{figure}

It is now simple to work out the matrix propagators. For the scalar field,
it is 
\be
\bD(k_0,\vec k)=\int_{-\infty}^{\infty}\frac{dk_0'}{2\pi}\rho(k_0',\vec k)
\bL (k_0',k_0)
\label{dlam}
\ee
with the elements of the matrix $\bL$ given by
\bea
&& \Lm_{11}=-\Lm_{22}^* =\frac{1}{k_0'-k_0-i\eta}+2\pi if(k_0')\de(k_0'-k_0)
\nonumber\\
&& \Lm_{12}=\Lm_{21}=2\pi ie^{\beta k_0'/2}f(k_0')\de(k_0'-k_0)
\eea
The corresponding results for the spinor propagator is
\be
\bS(p_0,\vec p)=\int_{-\infty}^{\infty}
\frac{dp_0'}{2\pi}\sg(p_0',\vp)\bom (p_0',p_0)
\label{slam}
\ee
where the matrix $\bom$ has the elements,
\bea  
&&\Om_{11}=-\Om_{22}^*=\frac{1}{p_0'-p_0-i\eta}-2\pi
i\wt{f}(p_0')\de(p_0'-p_0)\nonumber\\
&&\Om_{12}=-e^{\beta\mu}\Om_{21}=-2\pi ie^{\beta p_0'/2}\wt{f}(p_0')\de(p_0'-p_0)
\eea
Eqs.~(24,25) and (26,27) constitute our form for the spectral representation of any
bosonic and any fermionic two-point function respectively. The two matrices
$\bL$ and $\bom$ are symmetrical, being related by the 
replacement $f \leftrightarrow -\wt{f}$. These describe the common
propagation properties of all elementary (particle) and composite fields,
the dependence on spin, etc being contained in the respective spectral
function.

We also note that the distribution-like functions $f$ and $\wt{f}$ given by
Eq.~(7) which occur in Eq.(25) and (27) are analytic functions, in contrast
to the distribution functions themselves given by Eq.(8). In particular, the
matrices $\bL$ and $\bom$ do not contain any $\theta$-function in the
energy variables $k_0 (p_0)$, which are present in the conventional form of
the propagators in presence of chemical potential. Thus, as we show in Sec.5,
the use of the above spectral reprsentations avoids the difficulty
encountered in integration of loop energies over the conventional form of
propagators.

Recalling Eq.~(9), we can, of course, readily restate the matrix elements
given by Eqs.~(25, 27) in terms of the particle distribution functions themselves 
as
\bea
&& \Lm_{11}=-\Lm_{22}^* =\frac{1}{k_0'-k_0-i\eta\ep(k_0)}+2\pi in\ep(k_0)
\de(k_0'-k_0)\nonumber\\
&& \Lm_{12}=\Lm_{21}=2\pi i\sqrt{n(1+n)}\ep(k_0)\de(k_0'-k_0)\nonumber\\  
&&\Om_{11}=-\Om_{22}^*=\frac{1}{p_0'-p_0-i\eta\ep(p_0)}-2\pi
iN_1\ep(p_0)\de(p_0'-p_0)\nonumber\\
&&\Om_{12}=-e^{\beta\mu}\Om_{21}=-2\pi ie^{\beta \mu/2}N_2\ep(p_0)\de(p_0'-p_0)
\eea
where $N_1$ is defined by Eq.~(10) above and $N_2$ by
\be
N_2(p_0)=\sqrt{n_+(1-n_+)}\tht(p_0)-\sqrt{n_-(1-n_-)}\tht(-p_0)
\ee

Note that we choose primed variables $(k_0', p_0')$ for the functions 
multiplying the delta-functions in Eqs.~(25,27), while we
choose unprimed ones $(k_0, p_0)$ for such functions in Eq.~(28). The first
choice allows us to integrate over loop energies in a simple way (see
Eq.~(46) below). The second choice makes the matrices diagonalising the 
propagators independent of the integration variable in the spectral
representations (see Eqs.~(32,33) below). 

The matrices $\bL$ and $\bom$ as defined by Eq.~(28) can now be diagonalised 
respectively by $\bU$ and $\bV$,
\bea
&&\bU (k_0)=\left(\begin{array}{cc}\sqrt{1+n} & \sqrt{n}\\\sqrt{n} & \sqrt{1+n}
\end{array}\right),\\
&&\bV (p_0)=\left(\begin{array}{cc}N_2/\sqrt{N_1} &
-\sqrt{N_1}e^{\beta\mu/2}\\
 \sqrt{N_1}e^{-\beta\mu/2} & N_2/\sqrt{N_1}
\end{array}\right)
\eea
which in turn diagonalises the two-point correlation functions themselves,
\bea
&&\bD(k_0,\vk)=\bU (k_0)
\left(\begin{array}{cc}\oD & 0\\0 & -\oD^{\,*}\end{array}\right)\bU (k_0)\\
&&\bS(p_0,\vec p)=\bV (p_0)\left(\begin{array}{cc}\oS  & 0\\0 &
-\oS^{\,*}\end{array}\right) \bV (p_0),
\eea
where the analytic amplitudes $\oD$ and $\oS$ are given by
\bea
&&\oD (k_0,\vk)=\int_{-\infty}^{\infty}
\frac{dk_0'}{2\pi}\frac{\rho(k_0',\vk)}{k_0'-k_0-i\eta\ep(k_0)}\\ 
&&\oS (p_0,\vp)=\int_{-\infty}^{\infty}
\frac{dp_0'}{2\pi}\frac{\sg(p_0',\vp)}{p_0'-p_0-i\eta\ep(p_0)}
\eea 

To find the spectral densities $\rho$ and $\sg$, we need not refer to their
defining equations (15) and (20) any more. It suffices to calculate the
imaginary part of any one, say the $11$-component of the two point functions
to get them. Thus from Eqs.(25) and (27), we get
\bea
&&\rho (k_0,\vk)=2\tanh(\beta k_0/2) \mathrm{Im} D_{11} (k_0,\vk)\\   
&&\sg (p_0,\vp)=2\coth(\beta (p_0-\mu)/2) \mathrm{Im} S_{11} (p_0,\vp)
\eea

We close our discussion of the form of two-point functions by obtaining  
explicitly the conventional form of the free propagators for the scalar and 
the spinor fields \cite{Niemi,Kobes1}. With the free field spectral
densities given by Eqs.~(5, 6), Eqs. (34, 35) give immediately
\bea
&&\oD^{(0)} (k)=\frac{-1}{k^2-m_B^2+i\eta}\equiv \De(k,m_B)\nonumber\\
&&\oS^{(0)} (p)=(\ps+m_F)\De(p,m_F)
\eea
which are just the Feynman propagators in vacuum.
Then multiplying out the matrices in Eqs.(32, 33) we get the scalar propagator as 
\be
\bD^{(0)}(k_0,\vec k)=\left(\begin{array}{cc}d_{11} & d_{12}\\
d_{21} & d_{22}\end{array}\right)
\ee
where the matrix elements are
\bea
&&d_{11}=-d^*_{22}=\De(k,m_B)+2\pi in\de(k^2-m_B^2)\nonumber\\
&&d_{12}=d_{21}=2\pi i\sqrt{n(1+n)}\de(k^2-m_B^2)
\eea
and the spinor propagator as
\be
\bS^{(0)}(p_0,\vec p)=(\ps +m_F)\left(\begin{array}{cc}s_{11} & s_{12}\\
s_{21} & s_{22}\end{array}\right)
\ee
with the matrix elements
\bea
&&s_{11}=-s^*_{22}=\De(p,m_F)-2\pi iN_1\de(p^2-m_F^2)\nonumber\\
&&s_{12}=-e^{\beta\mu}s_{21}=-2\pi ie^{\beta
\mu/2}N_2\de(p^2-m_F^2)
\eea

\section{An example}

Finally we evaluate in some detail the energy integral in a loop involving a 
scalar (with $\mu=0$) and a spinor (with $\mu \neq 0$) propagator, both exact, 
using their spectral representations given by Eqs.(24-27). The matrix amplitude of 
such a loop, call it $\bsg$, belongs to the complete spinor two-point function and 
by Eqs.(35,37) its analytic amplitude $\overline{\Sg}$ must have the form
\be
\overline{\Sg} (p_0,\vec p)=\int_{-\infty}^{\infty}\frac{dp'_0}{2\pi}
\frac{\sg_1 (p_0',\vp)}{p_0'-p_0-i\eta\ep(p_0)}
\ee
where the one-loop spectral density is given by Eq.~(37) as
\be
\sg_1(p_0,\vp)=2\coth\{\beta(p_0-\mu)/2\}\mathrm{Im}\Sg_{11}(p_0,\vp).
\ee
 
Ignoring the spin and isospin structures of the interaction vertices, the
$11$-component of the matrix amplitude is given by
\bea
&&\Sg_{11}(p_0,\vp)=i\int\frac{d^4q}{(2\pi)^4}S_{11}(q)D_{11}(p-q)=
\int\frac{d^3q}{(2\pi)^3}\times\nonumber\\
&&\int_{-\infty}^{\infty}\frac{dq_0'}{2\pi}
\sg (q_0',\vec q) \int_{-\infty}^{\infty}\frac{dq''_0}{2\pi}
\rho (q''_0,\vec p-\vec q)\, I(q_0',q_0'',p_0)
\eea
where
\be
I(q_0',q''_0,p_0)=i\int_{-\infty}^{+\infty}\frac{dq_0}{2\pi}
\Om_{11}(q_0',q_0)\Lm_{11}(q''_0,p_0-q_0)
\ee
Rewriting the propagators in $I$ as
\bea
\Om_{11}(q_0',q_0)&=&\frac{1-\wt f(q_0')}{q_0'-q_0-i\eta}+\frac{\wt f(q_0')}
{q_0'-q_0+i\eta}
\nonumber\\
\Lm_{11}(q_0',q_0)&=&\frac{1+f(q_0')}{q_0'-q_0-i\eta}-\frac{f(q_0')}
{q_0'-q_0+i\eta}
\eea
we can integrate the resulting four terms over $q_0$ by just closing the contour in 
the $q_0$ plane to get 
\be 
I(q_0',q''_0,p_0)=\frac{\{1-\wt f(q_0')\}\{1+f(q''_0)\}}{p_0-q'_0-q''_0+i\eta}
+\frac{\wt f(q_0')f(q''_0)}{p_0-q'_0-q''_0-i\eta}
\ee
Had we taken the spinor propagator from Eq.(28), the $\tht$-functions in it
would not have allowed such a simple evaluation of $I$ \cite{Kobes2}. 

The imaginary part of $\Sg_{11}$ is given by that of $I$ 
\be
{\rm Im}I(q_0',q''_0,p_0)=-\pi \{(1-\wt{f})(1+f)-\wt{f}f)\}
\de(p_0-q_0'-q_0'')\,,\nonumber\\
\ee
which has the 'wrong' relative sign between the two terms in the bracket. It
can be reversed by taking out a factor of $\tanh \{\beta(p_0-\mu)/2\}$. Then
$F$ reduces to 
\bea
\sg_1(p_0,\vp)=&&-2\pi\int\frac{d^3q}{(2\pi)^3}\int_{-\infty}^{\infty}\frac{dq_0'}{2\pi}
\sg (q_0',\vec q) \int_{-\infty}^{\infty}\frac{dq''_0}{2\pi}\nonumber\\
&&\rho (q''_0,\vec p-\vec q)
\{1-\wt{f}(q_0')+f(q_0'')\}\de (p_0-q_0'-q_0'')\nonumber\\
\eea 

Further, if the propagators are the free ones, one can easily work out the
$q_0'$ and $q_0''$ integrals over the mass shell $\de$-functions in the 
spectral densities to get [ $\om_1=\sqrt{\vq^{~2}+m_F^2}\,,~~
\om_2=\sqrt{(\vp-\vq)^2+m_B^2}$ ]
\bea
&&\sg_1^{(0)}(p_0,\vp)=-2\pi\int\frac{d^3q}{(2\pi)^34\om_1\om_2}\left[(\gm_0\om_1-
\vec \gm\cdot \vec q +m_F)\right.\times\nonumber\\
&&\{(1-n_++n)\de(p_0-\om_1-\om_2)+(n_++n)\de(p_0-\om_1+\om_2)\}\nonumber\\
&& -(\om_1\rw -\om_1, \om_2\rw -\om_2, n_+\rw n_-)\left.\right]~,
\eea
in agreement with the computation in the imaginary time formulation \cite{Weldon}.

\section{Discussion}

Despite many applications of the real-time propagators in medium, their
conventional matrix form is not satisfactory in that the matrix components 
contain non-analytic functions, particularly when chemical potential
is present. This causes difficulty in carrying out integrals in loop
energies over such form of propagators.

In looking for improvement on this point, we take our cue from the
K\"{a}llen-Lehmann representation for the vacuum two-point functions, where
the propagation properties common to all fields are displayed explicitly by 
analytic functions, keeping dependence on spin etc. confined to the spectral 
densities. We are thus led to study this representation at finite temperature 
and chemical potential for all two-point correlation functions.

Indeed, we find spectral representations here to depend explicitly only on
analytic, distribution-like functions, along with propagation function expected 
in vacuum. They are only of two types, one for the bosonic and another for 
fermionic two-point functions. Also they are simple and symmetrical between the 
two types. We choose this representation even for free propagators. Then the loop 
integrals in the energy variables can be easily carried out, irrespective of 
the presence of chemical potentials in them. Integrals over the spectral densities 
can then be performed.

\end{document}